\documentclass[twocolumn,showpacs,prb]{revtex4}
\usepackage{amsfonts}
\usepackage{amsmath}
\usepackage{amssymb}
\usepackage{graphicx}
\usepackage{epstopdf}

\begin{document}

\title{Ferromagnetism and quantum anomalous Hall effect in
one-side-saturated buckled honeycomb lattices}
\author{Shin-Ming Huang$^{1}$, Shi-Ting Lee$^{1}$, and Chung-Yu Mou$%
^{1,2,3,4}$}
\affiliation{$^1$Department of Physics, National Tsing Hua University, Hsinchu 30043,
Taiwan\\
$^2$Frontier Research Center on Fundamental and Applied Sciences of Matters,
National Tsing Hua University, Hsinchu 30043, Taiwan\\
$^{3}$Institute of Physics, Academia Sinica, Nankang, Taiwan\\
$^{4}$Physics Division, National Center for Theoretical Sciences, P.O.Box
2-131, Hsinchu, Taiwan}

\begin{abstract}
The recently synthesized silicene as well as theoretically discussed
germanene are examples of buckled honeycomb structures. The buckled
structures allow one to manipulate asymmetry between two underlying
sublattices of honeycomb structures. Here by taking germanene as a prototype
of buckled honeycomb lattices, we explore magnetism induced by breaking
sublattice symmetry through saturating chemical bonds on one-side of the
buckled honeycomb lattice. It is shown that when fractions of chemical bonds
on one-side are saturated, two narrow bands always exist at half filling.
Furthermore, the narrow bands generally support flat band ferromagnetism in
the presence of the Hubbard $U$ interaction. The induced magnetization is
directly related to the saturation fraction and is thus controllable in
magnitude through the saturation fraction. Most importantly, we find that
depending on the saturation fraction, the ground state of an one-side
saturated germanene may become a quantum anomalous Hall (QAH) insulator
characterized by a Chern number that vanishes for larger magnetization. The
non-vanishing Chern number for smaller magnetization implies that the
associated quantum Hall effect tends to survive at high temperatures. Our
findings provide a potential method to engineer buckled honeycomb structures
into high-temperature QAH insulators.
\end{abstract}

\maketitle

\section{Introduction}

Since the pioneer work by Kane and Mele in 2005, \cite{Kane2005} the quantum
spin Hall (QSH) state has gathered great interest in the condensed matter
field. QSH state is characterized by an insulating bulk and gapless helical
edge states. \cite{Hasan2010,Konig2008} The original candidate with QSH
effect proposed by Kane and Mele \cite{Kane2005} was graphene in which
spin-orbit coupling (SOC) opens a gap at the Dirac points and turns the
semimetal into a topological insulator. However, SOC in graphene was
reported very weak such that QSH effect can be observed only at low
temperatures. \cite{Min2006,Yao2007} For stronger SOC, heavier elements are
possible candidates. Honeycomb lattice formed by other elements of group IV,
such as Si and Ge which are termed as silicene and germanene, respectively,
are predicted to be stable as planar lattices by first-principles
calculations. \cite{Takeda1994,Cahangiro2009,Liu2011a,Liu2011b} A prominent
feature of these materials is the buckled geometry, which allows a tunable
band gap via the vertical electric field. \cite{Ezawa2012a,Ni2012} Although
silicene or germanene has not been successfully isolated yet, recent
synthesis of silicene through epitaxial growth on silver substrate indicates
the feasibility of realizing silicene and germanene. \cite%
{Lalmi2010,Vogt2012,Lin2012} It is thus possible that silicene and
germanene could follow the graphene and open new perspectives for
applications, especially due to their compatibility with Si-based
electronics. % While so far measurements of the
%intrinsic properties are not feasible due to that the synthesized silicene is on
%conducting silver substrate and strong hyperdization with substrate may destroy intrinsic properties of silicene

In addition to the QSH state, another interesting state in graphene-related
materials that has been proposed theoretically is the quantum anomalous Hall
(QAH) state. \cite{Haldane1988,Onoda2003,Qiao2010,Tse2011,Ezawa2012b}
Similar to the quantum Hall state, the QAH state is also an insulating state
without time-reversal symmetry and is characterized by quantized Hall
conductance $C_{n}e^{2}/h$ ($C_{n}$, the Chern number) and the presence of
chiral edge states. However, unlike the quantum Hall state which originates
from quantized Landau levels induced by magnetic fields, QAH state arises by
nontrivial topology of electronic states associated with SOC and internal
magnetization. \cite{Qiao2010,Liu2008,CWu2008,Tse2011,Qiao2012} Some
proposals of the QAH effect on graphene have been suggested like via adatoms
\cite{HZhang2012} and proximity. \cite{Qiao2014} Due to the difficulty in
controlling magnetization and SOC, the QAH effect has not been observed
experimentally until very recently it is realized in a magnetically doped
topological insulator of Cr-doped (Bi,Sb)$_{2}$Te$_{3}$, \cite{Chang2013}
where the $C_{n}=1$ Hall conductance has been observed in low temperatures
around hundreds of mK. The realization of the QAH states has revived hopes
for using dissipationless edge states to develop low-power-consumption
electronics. However, from fundamental and practical points of view, QAH
states with larger band gaps and larger Chern numbers \cite%
{HJiang2012,Qiao2013,JWang2013,CFang2014} have the advantage of a lower
contact resistance and possible applications at higher temperatures.
Therefore, it is desirable to search for QAH insulators with larger Chern
numbers and larger band gaps. \cite{HJiang2012,Qiao2013,JWang2013,CFang2014}

In this paper, we explore QAH effects driven by spontaneous ferromagnetic
(FM) order induced by the Hubbard $U$ interaction in a buckled honeycomb. While
our theory and results apply to silicene as well, we shall take germanene as a 
prototype of buckled honeycomb lattices due to its larger spin-orbit coupling.
We show that instead of doping the system 
using magnetic atoms, \cite
{Chang2013} magnetism can be spontaneously induced by breaking sublattice
symmetry through saturating chemical bonds on one-side of the buckled
honeycomb lattice. 
%While our theory and conclusions apply to silicene as well, 
%The difference is just its smaller SOC couplings.)
One of the ways to
saturate the chemical bonds is to hydrogenate silicene or germanene. \cite
{Elias2009} Ferromagnetism in semi-hydrogenated honeycomb structures has
been reported by means of first-principles calculations for graphene \cite%
{JZhou2009} and silicene. \cite{XQWang2012} In our work, we find that
saturation of chemical bonds generally tends to localize electrons and
results in flat bands. Furthermore, in the presence of the Hubbard $U$
interaction, spontaneous ferromagnetism is generally induced in the flat
bands. The flat band ferromagnetism for the QAH effect is studied on the
dice lattice \cite{FWang2011} and a decorated lattice. \cite{Zhao2012} Here
we investigate the QAH effect in  flat bands of one-side-saturated
buckled honeycomb lattice for different saturation fractions characterized
by $1/q$ with $q$ being an integer. For $q=1,2,...,6,$ we find that the
critical $U$ for ferromagnetism increases with increasing $q$. The induced
magnetization is directly related to the saturation fraction and is thus
controllable in magnitude through the saturation fraction. Furthermore, we
find that QAH states with $C_{n}=-1$ or $2$ can be realized among these
saturation fractions. %There even exists a Chern number two phase in the
%latter system. The condition for the topological insulators is the intrinsic
%SOC ($\lambda _{SO}$) smaller than the ferromagnetic (FM) gap ($\Delta_{\rm{FM}}$%
%). Unfortunately, they are still not allowed to observe because the small FM
%gap has to be small. However, we give a clue to a small FM gap like by
%decreasing the vacancy concentration.
Specifically, the band gap of the system is given by the difference of the
intrinsic SOC ($\lambda _{SO}$) and the FM gap ($\Delta _{\mathrm{FM}}$).
For smaller magnetization, $\Delta _{\mathrm{FM}}<\lambda _{SO}$ and the
Chern number is nonzero. As a result, the Chern number $C_{n}$ starts from $%
-1$ for a smaller magnetization and becomes 0 when the magnetization is
large enough (a $C_{n}=2$ phase precedes the $C_{n}=-1$ phase in the
1/4-vacancy system). The non-vanishing Chern number for smaller magnetization
implies that the associated QAH effect tends to survive at high
temperatures. Our findings thus provide a potential method to engineer
buckled honeycomb structures into high-temperature QAH insulators.

This paper is organized as follows. In Section II, we present the
theoretical model of partially saturated germanene (or silicene) with
electron-electron interactions. Here the saturation of chemical bonds is
modeled as on-site impurity potentials on \textit{A} sites with strength $%
V_{A}$ . We will focus on the infinite potential limit so that sites with
saturated chemical bonds can be treated as vacancies. Section III is devoted
to the case of semi-saturation in which all \textit{A} sites are vacancies.
In Section IV, we investigate saturation fraction of \textit{A} sites being
1/2, 1/3 and 1/4. We summary our results including those of the 1/5- and
1/6-vacancy systems and give discussions in Sec. V.

\section{Theoretical Model}

We start by modeling silicene and germanene using the Kane-Mele model on a
honeycomb lattice. The honeycomb lattice for germanene and silicene is
different from that of graphene due to the buckled structure. Two
sublattices, labeled by \textit{A} and \textit{B}, are in different planes
shifted by some distance. In the absence of interactions and impurities, the
tight-binding model for both germanene and silicene is given by \cite%
{Liu2011b,Ezawa2012b}
\begin{eqnarray}
H\ &=&-t\sum_{\left\langle i,j\right\rangle \alpha }c_{i\alpha }^{\dag
}c_{j\alpha }+i\frac{\lambda _{SO}}{3\sqrt{3}}\sum_{\left\langle
\left\langle i,j\right\rangle \right\rangle \alpha \beta }\nu
_{ij}c_{i\alpha }^{\dag }\left( \hat{\sigma}_{z}\right) _{\alpha \beta
}c_{j\beta }  \notag \\
&&-i\frac{2}{3}\lambda _{R2}\sum_{\left\langle \left\langle i,j\right\rangle
\right\rangle \alpha \beta }\mu _{i}c_{i\alpha }^{\dag }\hat{z}\cdot \left(
\mathbf{\hat{\sigma}\times \hat{d}}_{ij}\right) _{\alpha \beta }c_{j\beta }
\label{hamiltonian}
\end{eqnarray}%
Here $\alpha $ and $\beta $ are indices for spin. The first term is the
nearest-neighbor hopping (between \textit{A} and \textit{B} sites). The
second term is the intrinsic SOC with $\nu _{ij}=(2/\sqrt{3})\hat{z}\cdot
\mathbf{\hat{d}}_{kj}\times \mathbf{\hat{d}}_{ik}=\pm 1$, where $\mathbf{%
\hat{d}}_{kj}$ and $\mathbf{\hat{d}}_{ik}$ are two unit vectors connecting
\textit{j} and \textit{i}. The third term is the next-nearest-neighbor
Rashba SOC where $\mu _{i}$ alternates sign between sublattices due to the
buckled structure. We will set the primitive lattice constant \textit{a} as
unity and adopt parameters for germanene, Following Liu \textit{et al}.'s
parameters, \cite{Liu2011b,Ezawa2012b} we shall neglect the nearest-neighbor
Rashba SOC due to its minute value: $t=1.3$ eV, $\lambda _{SO}=43$ meV, and $%
\lambda _{R2}=10.7$ meV. In this non-interacting system without external
fields, electrons are characterized by massive Dirac spectra at $K=(\frac{%
4\pi }{3},0)$ and $-K$ which give rise to the QSH effect.

To include the electron-electron interaction and characterize saturation of
chemical bonds on \textit{A} sites, additional terms with the on-site
potential $V_{i}$ and the Hubbard $U$ interaction are included as
\begin{equation}
\Delta H=\sum_{i\in A,\alpha }V_{i}c_{i\alpha }^{\dag }c_{i\alpha
}+U\sum_{i}c_{i\uparrow }^{\dag }c_{i\uparrow }c_{i\downarrow }^{\dag
}c_{i\downarrow }.
\end{equation}%
Here site $i$ is assigned a vacancy by setting $V_{i}$ to infinity if the
chemical bond of site $i$ is saturated, otherwise $V_{i}$ is set to zero. In
the mean-field approach, the on-site potential will be renormalized by
adding the following term
\begin{equation}
\Delta V=U\sum_{i}\sum_{\alpha }\left( n_{i}/2-\alpha m_{i}\right)
c_{i\alpha }^{\dag }c_{i\alpha },
\end{equation}%
where $n_{i}=\sum_{\alpha }\left\langle c_{i\alpha }^{\dag }c_{i\alpha
}\right\rangle $ and $m_{i}=\frac{1}{2}\sum_{\alpha }\alpha \left\langle
c_{i\alpha }^{\dag }c_{i\alpha }\right\rangle $ is the magnetic
polarization. In the followings except specified, units of the energies are
in terms of eV.

According to the work by Mielke and Tasaki, \cite{Mielke1991,Tasaki1998} for
a system with an isolated flat band or a nearly flat band system, a finite
Coulomb interaction may reach the Stoner criterion and generally results in
ferromagnetism in the flat band. However, for the \textit{p}-orbital
electronic system, the bandwidth is often too large so that the Coulomb
interaction is not strong enough to lift up the high degeneracy of ground
states. In the following, we shall show that for a bipartite system, the
asymmetry in two sublattice introduced by removing points in one sublattice
can generally induce magnetism. For bipartite lattices without SOC, the
induced magnetism results from the difference in number of lattice points in
two sublattices and agrees with the expectation from the Lieb's theorem.
\cite{Lieb} In the presence of spin-orbit interaction, removing points in
one sublattice generally results in a flat band due to suppression of the
nearest-neighbor hopping around vacancies. The emergence of an isolated flat
band enables the realization of magnetism in germanene and silicene.

\section{Full Saturation}

We start by considering the case in which a uniform potential $V_{A}$ is
applied at all \textit{A} sites. We will investigate finite potential at the
beginning and then focus on infinite potential. In the absence of
electron-electron interaction, the Hamiltonian at momentum $\mathbf{k}$ is
given by
\begin{equation}
\mathcal{H}=\frac{V_{A}}{2}\left( 1+\hat{\tau}_{z}\right) +T_{SO}\hat{\tau}%
_{z}\hat{\sigma}_{z}+\left( T\hat{\tau}_{+}+r\hat{\tau}_{z}\hat{\sigma}%
_{+}+H.c.\right) ,
\end{equation}%
where $\hat{\tau}$ and $\hat{\sigma}$ are the Pauli matrices in the $A-B$
sublattice space and the spin space, and
\begin{eqnarray}
T &=&-t\left[ e^{ik_{y}/\sqrt{3}}+2e^{-ik_{y}/2\sqrt{3}}\cos \left( \frac{1}{%
2}k_{x}\right) \right] , \\
T_{SO} &=&\frac{2\lambda _{SO}}{3\sqrt{3}}\left[ \sin \left( k_{x}\right)
-2\sin \left( \frac{1}{2}k_{x}\right) \cos \left( \frac{\sqrt{3}}{2}%
k_{y}\right) \right] , \\
r &=&-\frac{4}{3}\lambda _{R2}\left\{ \sqrt{3}\cos \left( \frac{1}{2}%
k_{x}\right) \sin \left( \frac{\sqrt{3}}{2}k_{y}\right) \right. \\
&&\left. +i\left[ \sin \left( k_{x}\right) +\sin \left( \frac{1}{2}%
k_{x}\right) \cos \left( \frac{\sqrt{3}}{2}k_{y}\right) \right] \right\} .
\notag
\end{eqnarray}%
The combination of different terms in spin results in an easy axis $%
z^{\prime }$ in \textbf{k} space so that the Hamiltonian can be rewritten as
\begin{equation}
\mathcal{H}=\frac{V_{A}}{2}+\left( \frac{V_{A}}{2}+\sqrt{T_{SO}^{2}+|r|^{2}}%
\hat{\sigma}_{z^{\prime }}\right) \hat{\tau}_{z}+T\hat{\tau}_{+}+T^{\ast }%
\hat{\tau}_{-}.
\end{equation}%
The four energy eigenvalues can be then explicitly found in the large $V_{A}$%
\ limit as
\begin{eqnarray}
E_{\sigma }^{U_{p}/L_{o}} &=&\frac{V_{A}}{2}\pm \sqrt{\left( \frac{V_{A}}{2}%
+\sigma \sqrt{T_{SO}^{2}+|r|^{2}}\right) ^{2}+|T|^{2}}  \notag \\
&\approx &\left\{
\begin{array}{ll}
V_{A}+\sigma \sqrt{T_{SO}{}^{2}+|r|^{2}}+\frac{|T|^{2}}{V_{A}} &  \\
-\sigma \sqrt{T_{SO}^{2}+|r|^{2}}-\frac{|T|^{2}}{V_{A}} &
\end{array}%
\right. ,
\end{eqnarray}%
where $U_{p}$ and $L_{o}$ denote upper and lower bands which are separated
by a gap of $V_{A}$ and each one has two sub-bands labeled by $\sigma =\pm $%
. In the limit of large $V_{A}$, the bandwidths for the upper $(U_{p})$ and
lower $(L_{o})$ bands are determined by SOC ($\lambda _{SO}$ and $\lambda
_{R2}$) instead of the hopping integral $t$. The minimum of $\sqrt{%
T_{SO}{}^{2}+|r|^{2}}$ is zero at $\Gamma $ and $M=(0,\frac{2\pi }{\sqrt{3}}%
) $ points, at which $|T|^{2}/V_{A}$ are $9t^{2}/V_{A}$ and $t^{2}/V_{A}$,
respectively. Therefore, it is a metal or a semimetal when the chemical
potential falls inside the two lower bands. In the FM state with a FM gap $%
\Delta _{\mathrm{FM}}$, the ground state is an insulating state when $%
2\Delta _{\mathrm{FM}}>8t^{2}/V_{A}$.

\begin{figure}[tbp]
\begin{center}
\includegraphics[height=1.9251in,width=3.5267in] {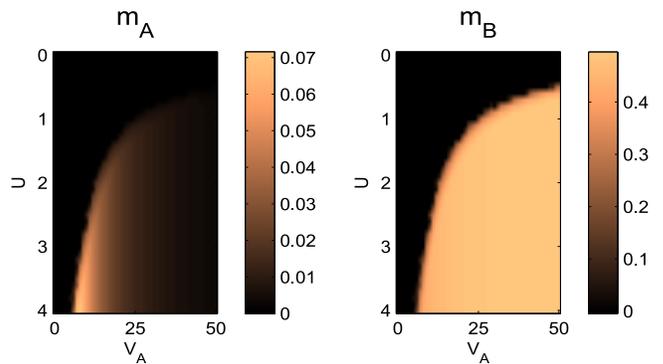}
\end{center}
\caption{(Color online) FM moments for finite $V_{A}$ at \textit{A} sites ($%
m_{A}$) and \textit{B} sites ($m_{B}$) when $n=n_{A}+n_{B}=1$.}
\label{Sz_VA_U}
\end{figure}

We first examine the phase diagram for ferromagnetism $V_{A}-U$ space for
the electron concentration $n=n_{A}+n_{B}=1$. In Fig. \ref{Sz_VA_U}, it is
shown that ferromagnetism happens at large $U$ and/or $V_{A}$, resulting
from a large ratio of interaction to bandwidth. Although both two $U$ and $%
V_A$ drive the system toward the strong coupling limit, there is a
difference between them: infinite $V_{A}$ does not lead to the flat band
limit but to a finite bandwidth limited by SOC (exact value being $2\lambda
_{SO}$). Therefore, a finite value of \textit{U} is required for a FM state.
Figure \ref{Sz_VA_U} also shows that the moment at \textit{B} sites ($m_{B}$%
) quickly saturates when entering the FM phase, while the moment at \textit{A%
} sites ($m_{A}$) gradually decreases as $V_{A}$ increases.

\begin{figure}[tbp]
\begin{center}
\includegraphics[height=1.92in,width=3.401in] {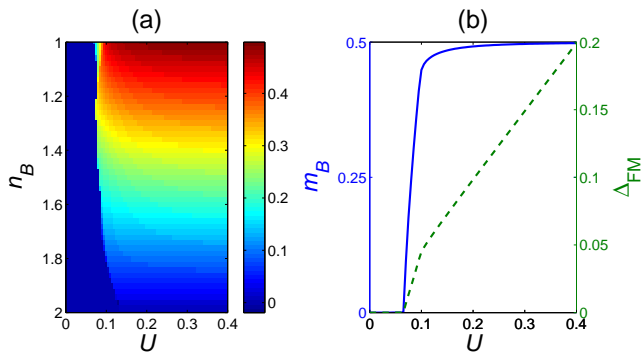}
\end{center}
\caption{(Color online) FM moment when all \textit{A} sites are saturated by
infinite $V_A$. Here the magnetic moment only survives at \textit{B} sites
with $m_{B}\equiv \left( n_{B\uparrow }-n_{B\downarrow }\right) /2$. (a)
shows the dependence of magnetic momentum on $n_{B}$ and $U$, while $m_B$
versus $U$ for $n_{B}=1$ is shown in (b) (blue solid line). The FM gap $%
\Delta_{\mathrm{FM}}=Um_{B}$ is also shown in panel (b) by the green dashed
line.}
\label{mB_uniform}
\end{figure}

In the following we will focus on the limit, $V_{A}\rightarrow \infty $. In
this limit, the self-consistent equation for the magnetic moment at \textit{B%
} site is
\begin{equation}
m_{B}=\frac{1}{N}\sum_{\mathbf{k}}\frac{T_{SO}^{\prime }(\mathbf{k})}{2E_{g}(%
\mathbf{k})}\left[ n_{-}(\mathbf{k})-n_{+}(\mathbf{k})\right] ,
\label{selfcons}
\end{equation}%
where $T_{SO}^{\prime }=T_{SO}+Um_{B}$, $E_{g}=\sqrt{\left( T_{SO}^{^{\prime
}}\right) ^{2}+|r|^{2}}$ and $n_{\pm }=\left[ \exp ((\pm E_{g}-\mu )/kT)+1%
\right] ^{-1}$. At zero temperature and $n_{B}=1$, the critical \textit{U}
is determined by
\begin{equation}
\frac{1}{U_{c}}=\frac{1}{N}\sum_{\mathbf{k}}\frac{1}{2E_{g}^{0}(\mathbf{k})},
\label{Uc}
\end{equation}%
where $E_{g}^{0}(\mathbf{k})$ is $E_{g}(\mathbf{k})$ for $m_{B}=0$. $T_{SO}$
in the numerator is absent because its sign oscillates and cancels exactly.
In the low energy region, the Rashba term is linear around $\Gamma $ and $M$%
, $r(\mathbf{k})\sim \lambda _{R2}(k_{y}+ik_{x})$, while $T_{SO}$ is
quadratic around $\Gamma $ point and linear around $M$ point, so a linear
dispersion displays. As a result, in two dimensions, the critical \textit{U}
is $U_{c}=c\lambda _{R2}$, where \textit{c} is a non-universal number
inversely proportional to the momentum cutoff. In Fig. \ref{mB_uniform}, we
show numerical solutions to Eq. (\ref{selfcons}). Left panel shows the
dependence of magnetic moment $m_{B}$ on $n_{B}$ and the interaction
strength \textit{U}, while the right panel shows $m_{B}$ (solid line) and
its corresponding gap $\Delta_{\mathrm{FM}}\equiv Um_{B}$ (dashed line) when
$n_B=1$.

Let us elucidate the symmetry of the FM state: the "inversion" symmetry ($%
\mathbf{k\rightarrow -k}$) is broken, while the threefold rotational
symmetry preserves. Although the lattice has $C_{6}$ symmetry, due to SOC,
the presence of ferromagnetism breaks the $C_{6}$ symmetry and therefore the
inversion symmetry is broken. Specifically, a $\pi /3$ rotation of \textbf{k}
results in $T_{SO}(\mathbf{k})\rightarrow -T_{SO}(\mathbf{k})$ and $r(%
\mathbf{k})\rightarrow r(\mathbf{k})e^{i\frac{\pi }{3}}$. Therefore, a
twofold $\pi /3$ rotation is a $2\pi /3$ rotation that keeps $T_{SO}^{\prime
}=T_{SO}-\Delta_{\mathrm{FM}}$ invariant, while a threefold $\pi /3$
rotation is an inversion and can not keep $T_{SO}^{\prime }=T_{SO}-\Delta_{%
\mathrm{FM}}$ invariant. As a result, the FM state breaks time-reversal and
inversion symmetries and gives rise to odd numbers of Fermi surfaces when it
is doped. In addition, the above symmetry property also requires the FM
moment being perpendicular to the \textit{x--y} plane; otherwise, the $C_{3}$
symmetry will be further lost due to the Rashba term.

We distract shortly to the case when the Rashba coupling is turned off. The
dominant low energy dispersion is quadratic now and hence the integral in
Eq. (\ref{Uc}) gives a logarithmic divergence. Instead of being finite, the
critical \textit{U} becomes zero. Hence the ground state of the system is
always FM for any non-vanishing $U$. In general, for quadratic band crossing,
other symmetry breaking might happen. \cite{KSun2009} However, for bipartite
lattices without SOC, according to the Lieb's theorem, finite magnetization
is induced by the difference of number of lattice points in two sublattices.
Results based on Eq. (\ref{Uc}) thus agrees with the expectation of the
Lieb's theorem.

We now examine the topology of the FM insulating state found in the above.
In the limit $V_{A}\rightarrow \infty $, the Hamiltonian contains only
\textit{B}-site electrons and can be expressed in terms of Pauli matrices $%
\hat{\sigma _{i}}$s , $H_{\mathbf{k}}^{B}\equiv E_{g}(\mathbf{k})\hat{d}(%
\mathbf{k})\cdot \vec{\sigma}$. By defining $\hat{d}(\mathbf{k})=(\sin
\theta _{\mathbf{k}}\cos \phi _{\mathbf{k}},-\sin \theta _{\mathbf{k}}\sin
\phi _{\mathbf{k}},\cos \theta _{\mathbf{k}})$, the Chern number specifying
quantum Hall states can be expressed by%
\begin{equation}
C_{n}=\frac{1}{4\pi }\int d^{2}\mathbf{k}\epsilon _{ij}\partial _{k_{i}}\cos
\theta _{\mathbf{k}}\partial _{k_{j}}\phi _{\mathbf{k}},  \label{Chern}
\end{equation}%
which counts the number of field space $\left( \theta _{\mathbf{k}},\phi _{%
\mathbf{k}}\right) $ covering a torus in the Brillouin zone (BZ). The
nontrivial topological phase is when $\phi _{\mathbf{k}}$ goes through $%
[-\pi ,\pi ]$ and $\theta _{\mathbf{k}}$ goes through $[0,\pi ]$ as \textbf{k%
} runs in BZ. Following Fukui and Hatsugai, \cite{Fukui} we compute the
Chern number numerically. The computed Chern number is shown in Fig. \ref%
{Cn_Bonly}. It is seen that the Chern number $C_{n}$ changes from $-1$ (QAH
insulator) to 0 (a trivial band insulator) as $\Delta _{\mathrm{FM}}$
increases, The topological phase transition is clearly driven by
magnetization. However, unlike the usual transition, nontrivial topology
disappears at larger magnetic moments. The topological phase exists when $%
\max [T_{SO}^{\prime }(\mathbf{k)]\times }\min [T_{SO}^{\prime }(\mathbf{k)]}%
<0$, that is $\Delta _{\mathrm{FM}}<\max [T_{SO}(\mathbf{k)]}=\lambda _{SO}$%
. At the critical point $\Delta _{\mathrm{FM}}=\lambda _{SO}$, the band gap
closes at $K$ point. Using Fig. \ref{mB_uniform}, one can estimate the
criterion on $U$ for the emergence of the QAH insulator. Since larger $U$
implies larger magnetic moment, results in Fig. \ref{mB_uniform} imply that
\textit{U} has to lie in the range $\text{0.07 eV}<U<\text{0.1 eV}$ in order
to observe the phase of the QAH insulator.
\begin{figure}[tbp]
\begin{center}
\includegraphics[height=2.34in,width=3.28in] {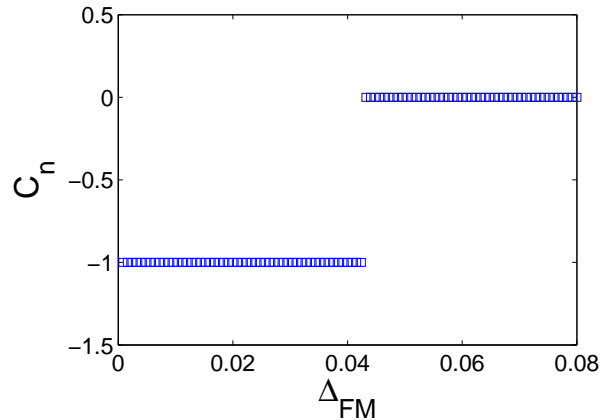}
\end{center}
\caption{(Color online) Chern number $C_{n}$ versus the FM gap $\Delta _{%
\mathrm{FM}}$ at $n_{B}=1$ when all chemical bonds on \textit{A} sites are
saturated. A topological phase transition occurs at $\Delta _{\mathrm{FM}}=%
\protect\lambda _{SO}=43$ meV.}
\label{Cn_Bonly}
\end{figure}

\section{Finite Saturation Fraction}

To explore other possible phases, we investigate situations when finite
fractions of \textit{A} sites are saturated. To saturate chemical bonds on
part of \textit{A} sites, we set $V_{A}$ to infinity at those \textit{A}
sites. Effectively, vacancies are introduced at those \textit{A} sites whose
chemical bonds are saturated.

In order to systematically investigate finite saturation fractions, we shall
consider situations with vacancies arranged in periodic patterns. As a
result of including vacancies in periodic patterns, the unit cell changes
and becomes a supercell. In the limit of $V_A \rightarrow \infty$, one needs
to consider lattice points in a supercell that are not vacancies. Therefore,
we shall consider a supercell in which number of lattice points that
excludes vacancies is $N_{c}$. After the Fourier transformation, the
Hamiltonian in Eq. (\ref{hamiltonian}) becomes
\begin{equation}
H = \sum_{\mathbf{k}} \Psi^{\dagger}_{\mathbf{k}} \mathcal{H}_{\mathbf{k}}
\Psi_{\mathbf{k}},
\end{equation}
where $\Psi _{\mathbf{k}}= \left(
\begin{array}{cc}
\Psi _{\uparrow }(\mathbf{k}) & \Psi _{\downarrow } (\mathbf{k})%
\end{array}
\right) ^{t}$ with $\Psi _{\sigma = \uparrow ,\downarrow } (\mathbf{k}) $
being a $N_c$-component electronic operator, $(c_{1 \sigma} (\mathbf{k}),
c_{2 \sigma} (\mathbf{k}), \cdots, c_{N_c\sigma}(\mathbf{k}) )^{t}$. $%
\mathcal{H}_{\mathbf{k}}$ is the Hamiltonian matrix and can be expressed in
the following block form
\begin{equation}
\mathcal{H}_{\mathbf{k}}=\left(
\begin{array}{cc}
K_{\mathbf{k}}+M_{\mathbf{k}} & \Delta_{\mathbf{k}} \\
\Delta_{\mathbf{k}}^{\dag } & K_{\mathbf{k}}-M_{\mathbf{k}}%
\end{array}
\right).  \label{H_su2}
\end{equation}
Here $K_{\mathbf{k}}$, $M_{\mathbf{k}}$, and $\Delta_{\mathbf{k}}$ are $N_c
\times N_c$ matrices. Furthermore, $K_{\mathbf{k}}$ includes the hopping
terms, $M_{\mathbf{k}}$ includes both the intrinsic SOC and the FM field,
and $\Delta_{\mathbf{k}}$ is the Rashba term.

We first note that there is a chiral symmetry in the matrix $K_{\mathbf{k}}$%
. Since there is no on-site energy and hopping only occur between nearest
neighbors (\textit{A} and \textit{B} sites), the main diagonal block of $K_{%
\mathbf{k}}$ vanishes. Therefore, the unitary matrix $U_l$ that changes $%
c_{A}\rightarrow c_{A}$ and $c_{B}\rightarrow -c_{B}$ transforms $K_{\mathbf{%
k}}$ into $U_{l}^{\dag }K_{\mathbf{k}} U_{l}=-K_{\mathbf{k}}$. However,
since both $M_{\mathbf{k}}$ and $\Delta_{\mathbf{k}}$ only couple \textit{A}
sites to \textit{A} sites or \textit{B} sites to \textit{B} sites, the
transformation $U_l$ leaves $M_{\mathbf{k}}$ and $\Delta_{\mathbf{k}}$
unchanged.

The chiral symmetry of $K_{\mathbf{k}}$ implies that for given an eigenstate
of $K_{\mathbf{k}}$, $\left\vert n\right\rangle $ with energy $\varepsilon
_{n}$, there must be another state $\left\vert \overline{n}\right\rangle
\equiv \left\vert -n\right\rangle =U_{l}\left\vert n\right\rangle $ with
negative energy $-\varepsilon _{n}$. Because $K_{\mathbf{k}}$ has $N_{c}$
eigenvalues, if $N_{c}$ is odd, there will be at least one zero mode for
each $K_{\mathbf{k}}$ block ($\sigma =\uparrow ,\downarrow $). Therefore,
without including effects due to intrinsic SOC ($M_{\mathbf{k}}$) and the
Rashba interaction ($\Delta _{\mathbf{k}}$), there are two energy bands
which are degenerate at zero energy. We shall denote states of the zero mode
by $|0,\mathbf{k},\sigma =\uparrow ,\downarrow \rangle $, if they are the
only eigenstates of $K_{\mathbf{k}}$, and term these two bands as midgap
bands. As shown for typical density of states for odd $N_{c}$ in Fig. \ref%
{dos}, the midgap bands persist even if $M_{\mathbf{k}}$ and $\Delta _{%
\mathbf{k}}$ are included.

%Furthermore, since
%all of vacancies are on $A$ sites,
%the supercell contains only $B$ sites. Hence the zero mode $| 0, \vec{k}, \sigma \rangle$ vanishes at
%$A$ sites and are nonvanishing only at $B$ site. As a result, since both $M_k$ and $\Delta_k$ do not
% couple $A$ and $B$ sites, there is no matrix element of  between $\left\vert 0\right\rangle $ and other $%
% \left\vert n\right\rangle $ states, \textit{i.e.}, $\left\vert
% 0\right\rangle $ is independent of others. Now we can construct the middle
% band in the basis of $\left\vert 0,\sigma = \uparrow ,\downarrow
% \right\rangle $. The kernel of the middle band will be simply a $2\times 2$
The emergence of midgap bands is similar to the arising of the impurity
bands due to random vacancies in graphene. \cite{mou} Without including
effects due to intrinsic SOC ($M_{\mathbf{k}}$) and the Rashba interaction ($%
\Delta _{\mathbf{k}}$), two energy bands are degenerate and are composed
only by wavefunctions at \textit{B} sites. Using $|0,\mathbf{k},\sigma
=\uparrow ,\downarrow \rangle $, the effective Hamiltonian for midgap bands
that include $M_{\mathbf{k}}$ and $\Delta _{\mathbf{k}}$ can be expressed by
a simple $2\times 2$ matrix
\begin{equation}
H_{mid}(\mathbf{k})=\left(
\begin{array}{cc}
m_{\mathbf{k}} & \delta _{\mathbf{k}} \\
\delta _{\mathbf{k}}^{\ast } & -m_{\mathbf{k}}%
\end{array}%
\right) ,  \label{Hmid}
\end{equation}%
where $m_{\mathbf{k}}=\langle 0,\mathbf{k},\uparrow |M_{\mathbf{k}}|0,%
\mathbf{k},\uparrow \rangle $ and $\delta _{\mathbf{k}}=\langle 0,\mathbf{k}%
,\uparrow |M_{\mathbf{k}}|0,\mathbf{k},\downarrow \rangle $. The energy
eigenvalues $E_{mid}^{\pm }(\mathbf{k})$ and eigenstates $|\pm ,\mathbf{k}%
\rangle $ for midgap bands can be estimated by diagonalizing $H_{mid}$ and
are given by $E_{mid}^{\pm }(\mathbf{k})=\pm E(\mathbf{k})$ with $E(\mathbf{k%
})=\sqrt{m_{\mathbf{k}}^{2}+\left\vert \delta _{\mathbf{k}}\right\vert ^{2}}$
and
\begin{equation}
\left(
\begin{array}{c}
|+,\mathbf{k}\rangle \\
|-,\mathbf{k}\rangle%
\end{array}%
\right) =\left(
\begin{array}{cc}
u_{\mathbf{k}} & v_{\mathbf{k}}e^{-i\phi _{\mathbf{k}}} \\
-v_{\mathbf{k}}e^{i\phi _{\mathbf{k}}} & u_{\mathbf{k}}%
\end{array}%
\right) \left(
\begin{array}{c}
|0,\mathbf{k},\uparrow \rangle \\
|0,\mathbf{k},\downarrow \rangle%
\end{array}%
\right) ,
\end{equation}%
where $u_{\mathbf{k}}=\sqrt{\frac{1}{2}(1+m_{\mathbf{k}}/E_{\mathbf{k}})}$,$%
v_{\mathbf{k}}=\sqrt{\frac{1}{2}(1-m_{\mathbf{k}}/E_{\mathbf{k}})}$, and $%
\phi _{\mathbf{k}}=\tan ^{-1}(\Im \delta _{\mathbf{k}}/\Re \delta _{\mathbf{k%
}})$. The magnetic moments in the following cases are found by solving Eq. (%
\ref{selfcons}) with $E_{g}$ being replaced by $E(\mathbf{k})$ and $%
T_{SO}^{\prime }$ by $-m_{\mathbf{k}}$.

\begin{figure}[tbp]
\begin{center}
\includegraphics[height=4.04in,width=2.18in] {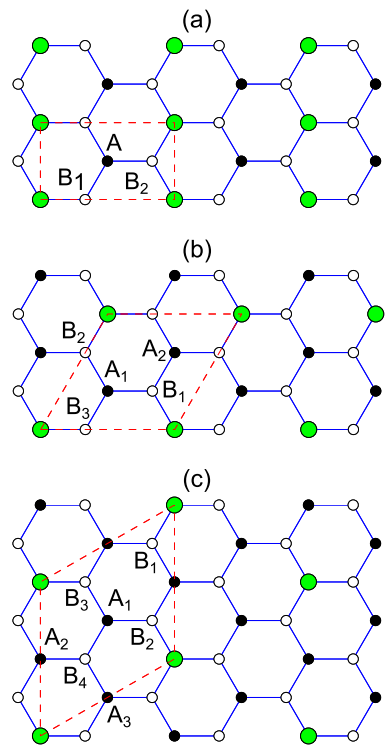}
\end{center}
\caption{(Color online) Vacancy arrangements for different vacancy
fractions: (a) 1/2-vacancy, (b) 1/3-vacancy, and (c) 1/4-vacancy. Here black
circles denote \textit{A} sites, empty circles denote \textit{B} sites, and
larger green circles denote vacancies. A unit cell (supercell) is enclosed
by dashed lines. The horizontal (vertical) direction is taken as the \textit{%
x} (\textit{y}) axis.}
\label{lattice}
\end{figure}
The vacancy configurations that we shall be focusing on are vacancies
arranged in periodic patterns. In general, there are \textit{p} vacancies in
a supercell with the original number of \textit{A} sites being \textit{q}.
Such configurations are dubbed as a \textit{p/q}-vacancy system with \textit{%
p, q} being mutually prime and $p<q$. In this configuration, a supercell
contains $q-p$ \textit{A}-site and \textit{q} \textit{B}-site atoms and
hence we have $N_{c}=2q-p$ and $2N_{c}$ bands. For each of band in the $%
2N_{c}$\ bands, the maximal filling is one electron per supercell. Hence
there will be maximally one electron in a unit cell belonging to a
particular band. However, for the midgap bands, since the wavefunction is
non-vanishing only at \textit{B} sites, the midgap bands are capable of
holding $2/q$ electrons per \textit{B} site. Hence at half filling, the
midgap bands contain $1/q$ electrons per \textit{B} site. We note that the
above argument is based on the assumption that $|0,\mathbf{k},\sigma
=\uparrow ,\downarrow \rangle $ are the only eigenstates of $K_{\mathbf{k}}$%
; if not, the midgap states will contain \textit{A}-site electrons as the
1/3-vacancy case in Sec. IV.B.

In this section, we will demonstrate three vacancy fractions: 1/2, 1/3 and
1/4-vacancy fractions. The vacancy patterns are illustrated in Fig. \ref%
{lattice}: (a) for 1/2-vacancy, (b) for 1/3-vacancy, and (c) for
1/4-vacancy. We also calculate 1/5- and 1/6-vacancy cases but to shorten the
context we only show their relevant results in the summary, Sec. V.
\begin{figure}[tbp]
\begin{center}
\includegraphics[height=4.00in,width=3.54in] {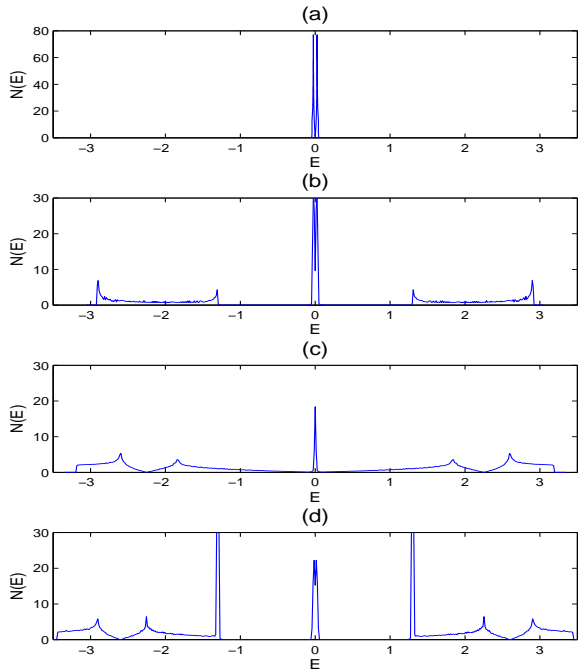}
\end{center}
\caption{(Color online) Density of states \textit{N}(\textit{E}) for the full saturated system (a), the
1/2-vacancy system (b), the 1/3-vacancy system (c), and the 1/4-vacancy
system (d)  at \textit{U} = 0 and half-filling. The energy \textit{E} is in the
unit of eV. In addition to midgap bands at center, there are extra side bands: two side bands in (b), four in
(c), and six in (d) (here the counting of bands does not include spin degeneracy). }
\label{dos}
\end{figure}

\subsection{1/2-vacancy}

\begin{figure}[tbp]
\begin{center}
\includegraphics[width=3.28in] {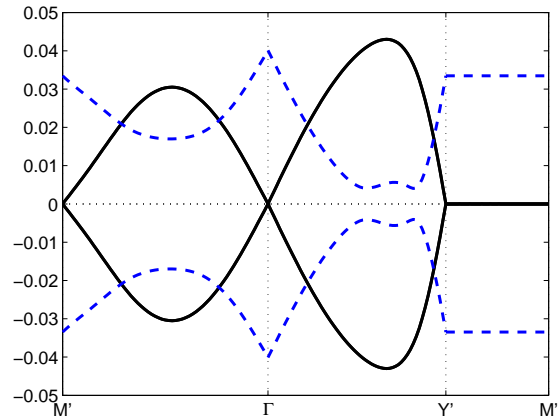}
\end{center}
\caption{(Color online) Band structures of the midgap bands in the
1/2-vacancy system: black solid lines for $U=0$ and dashed lines for $%
U=0.18$ eV ($\Delta _{\mathrm{FM}}=38$ meV). Since the inversion symmetry
is broken by ferromagnetism, band energies at $\mathbf{k}$ and $-\mathbf{k}$
are different. Here \textbf{k} is plotted from $\Gamma$ to $Y^{\prime }=-(0,\frac{\protect%
\pi }{a})$ or $M^{\prime }=-(\frac{\protect\pi }{\protect\sqrt{3}a},\frac{%
\protect\pi }{a})$ to exhibit the trend of band touching with increasing $\Delta _{%
\mathrm{FM}}$. The midgap bands touch right at $\mathbf{k}=-(0,\frac{2\protect\pi }{%
3a})$ when $\Delta _{\mathrm{FM}}=\protect\lambda _{SO}$.}
\label{Ek_one_half}
\end{figure}

The vacancy arrangement for 1/2-vacancy is illustrated in Fig. \ref{lattice}%
(a), in which a unit cell contains three atoms, $A$, $B_{1}$, and $B_{2}$.
Detailed band structures are presented in Appendix A1. In Fig. \ref{dos}(b),
we show the density of states when $U=0$. There are three bands (with
twofold degeneracy for each band): a narrow band and two wide bands,
separated by gaps about the order of $t$. As explained before, the midgap
bands and their bandwidths are solely determined by the SOC strength. Since $%
\lambda _{SO}$ is small, the midgap bands are almost flat and support
ferromagnetism. The band structures of the midgap bands are shown in
Fig. \ref{Ek_one_half} by the solid lines for $U=0$ and the dashed lines for $U \neq 0$.

We now examine the FM state in Fig. \ref{mz}(a) and (b). Fig. \ref{mz}(a)
shows the density plot of the average moment at \textit{B} sites $%
m_{B}=\left( m_{B1}+m_{B2}\right) /2$ versus $U$ and the average particle
density $n=\left( n_{A}+n_{B1}+n_{B2}\right) /3$. It is seen that for
moderate magnitude of $U$, ferromagnetism is always induced. At half
filling, the lowest band is completely filled and the chemical potential is
right at the center of the midgap band. As a result, the midgap bands make a
large contribution to magnetization. This is shown by the solid line in Fig. %
\ref{mz}(b). According to the previous elaboration, there is 1/2 electron
per $B$ site from each midgap band. Hence the maximal $m_{B}$ is 1/4, which
is in consistent with results shown in Fig. \ref{mz}(b). Note that in our
calculations, a tiny $m_{A}$ is present due to small deviation of $n_{A}$
and $n_{B}$ from one. In Fig. \ref{mz}(b), we also examine the FM gap $%
\Delta _{\mathrm{FM}}$ (the green dashed line) defined by $\Delta _{\mathrm{%
FM}}=Um_{B}$. Compared with the full-vacancy case in Fig. \ref{mB_uniform},
although $m_{B}$ saturates quickly also for 1/2-vacancy, both the
magnetization and $\Delta _{\mathrm{FM}}$ are weaker in all doping regime.

Finally, the topology of the 1/2-vacancy system at half filling is examined
by computing the Chern number. We find the same result as in Fig. \ref%
{Cn_Bonly} that the topologically nontrivial state ($C_{n}=-1$) exists for $%
\Delta _{\mathrm{FM}}<\lambda _{SO}$. The value of $U$ for $\Delta _{\mathrm{%
FM}}=\lambda _{SO}$ is about 0.195 eV. At the transition point, the midgap band 
touch at $\mathbf{k}=\left( 0,-\frac{2\pi }{3a}\right) $ as one can see
the tendency from the dashed lines shown in Fig. \ref{Ek_one_half}, which is for $%
\Delta _{\mathrm{FM}}=38$ meV before the transition.

\subsection{1/3-vacancy}

\begin{figure}[tbp]
\begin{center}
\includegraphics[width=3.40in] {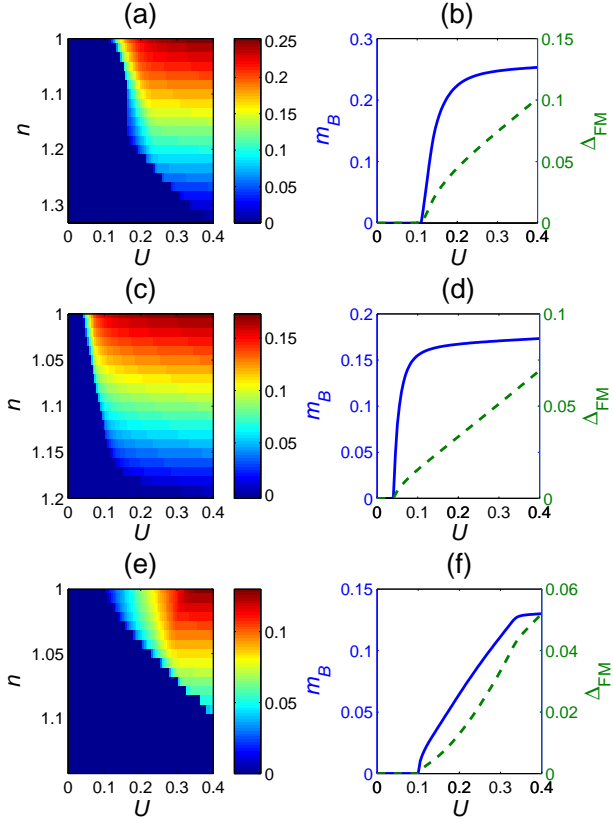}
\end{center}
\caption{(Color online) FM moment for the 1/2-vacancy [(a), (b)],
1/3-vacancy [(c), (d)], and 1/4-vacancy [(e), (f)] systems. Left panels (a),
(c), and (e) are for average magnetic moment $m_{B}$ on \textit{B} sites $%
m_{B}$ in the $n-U$ plane. Right panels (b), (d) and (f) show $m_B$ (blue
solid lines) and $\Delta _{\mathrm{FM}}$ (green dashed lines) versus $U$ at $n=1$. Here 
$\Delta_{\mathrm{FM}}=U m_B$.}
\label{mz}
\end{figure}
\begin{figure}[tbp]
\begin{center}
\includegraphics[width=3.28in] {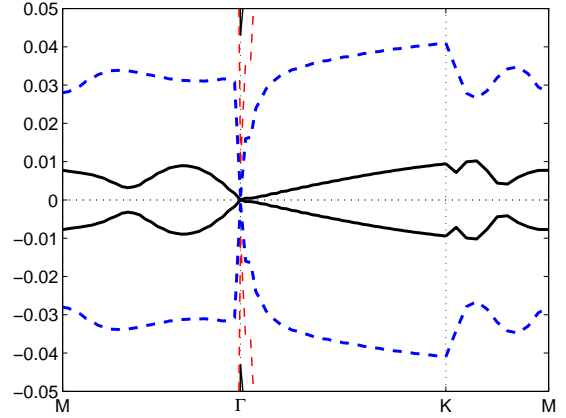}
\end{center}
\caption{(Color online) Band structures of midgap bands in the 1/3-vacancy
system: black solid lines for $U=0$ and blue dashed lines for $U=0.2$ eV ($%
\Delta _{\mathrm{FM}}=33$ meV). Band edges of two higher-energy bands for $
U=0.2$ eV are also show by red dashed lines, which move down to zero energy
at $\Gamma $point when the topological transition occurs.}
\label{Ek_one_third}
\end{figure}

As shown in Fig. \ref{lattice}(b), the unit cell for 1/3-vacancy lattice
contains five atoms, $A_{1}$, $A_{2}$, $B_{1}$, $B_{2}$, and $B_{3}$. The
detail of the bulk band structure is summarized in Appendix A2. The density
of states for $U=0$ is shown in Fig. \ref{dos}(c). There are five bands with
twofold degeneracy for each band. Apparently, there is no large gap opened
and the gaps between the midgap bands and its neighbors is $\lambda _{SO}$.
The band touches in Fig. \ref{dos}(c) are a numerical artifact. The fact
that gaps are of order of $\lambda _{SO}$ instead of $t$ is because there
are three zero-energy states of $K_{\mathbf{k}=\Gamma }$. One of three is a
state of \textit{A}-site electrons. As a result, the midgap states when SOC
turns on will not contain \textit{B}-site electrons only. However, we check
numerically that the portion of \textit{A}-site electrons is still tiny. The
band structures of the midgap bands are shown in Fig. \ref{Ek_one_third}
by the solid lines.

Characterization of the FM states are shown in Fig. \ref{mz}(c) and (d). Due
to particular symmetry in the lattice, the spin (and charge) density at
\textit{A} sites and \textit{B} sites are the same individually.
Furthermore, for the same reason as that for the 1/2-vacancy system, there
is 1/3 electron per \textit{B} site from each midgap band. Hence the maximal
$m_{B}$ is 1/6, which is weaker than that of the 1/2-vacancy system. In
addition, from the phase diagram in Fig. \ref{mz}(c) and (d), we find that
the critical $U$ for emergence of the FM state is smaller than that of the
1/2-vacancy system. This can be understood by the narrow bandwidth in Fig. %
\ref{Ek_one_third}. We show the energy dispersions for $\Delta _{\mathrm{FM}}=33$
meV by the dashed lines in Fig. \ref{Ek_one_third}: from the thick dashed
lines there is a tiny gap at $\Gamma $ due to the FM gap from \textit{A}%
-site electrons and high-energy bands will move toward zero energy and touch
at $\Gamma $ as indicated by the thin dashed lines.

The topology of the 1/3-vacancy system is examined at half filling. We find
the same result as cases before that the topologically nontrivial state ($%
C_{n}=-1$) exists for $\Delta _{\mathrm{FM}}<\lambda _{SO}$. The value of $U$
for $\Delta _{\mathrm{FM}}=\lambda _{SO}$ is about 0.255 eV.

\subsection{1/4-vacancy}

Our last example is the 1/4-vacancy system. As shown in Fig. \ref{lattice}%
(c), the unit cell for the lattice of the 1/4-vacancy system contains three
\textit{A} sites ($A_{1}$, $A_{2}$, and $A_{3}$) and four \textit{B} sites ($%
B_{1}$, $B_{2}$, $B_{3}$, and $B_{4}$). The detailed band structures are
summarized in Appendix A3. In Fig. \ref{dos}(d), we show the density of
states for $U=0$. It is clear that seven bands with twofold degeneracy for
each band can be recognized. The feature of the 1/4-vacancy system is
similar to that of the 1/2-vacancy system: there is a large energy gap of
order $t$. Band structures of the midgap bands can be seen in Fig. \ref%
{Ek_one_fourth} by the solid lines.
\begin{figure}[tbp]
\begin{center}
\includegraphics[width=3.28in] {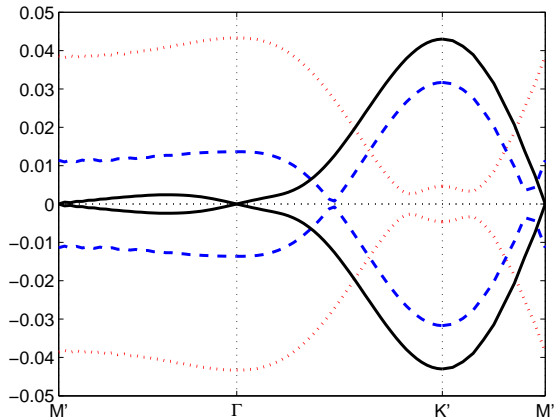}
\end{center}
\caption{(Color online) 
Band structures of the midgap bands in the 1/4-vacancy
system: black solid lines for $U=0$, blue dashed lines for $U=0.19$ eV ($%
\Delta _{\mathrm{FM}}=11$ meV), red dotted lines for $U=0.32$ eV ($\Delta _{%
\mathrm{FM}}=38$ meV). Note that \textbf{k} is plotted from $\Gamma$ to $K^{\prime }=-(0,%
\frac{{4\protect\pi }}{3c})$ or $M^{\prime }=-(\frac{{\protect\pi }}{{%
\protect\sqrt{3}c}},\frac{{\protect\pi }}{{c}})$ (${c=2a}$) in order to see
the trend of band touching with increasing $\Delta _{\mathrm{FM}}$. }
\label{Ek_one_fourth}
\end{figure}
\begin{figure}[tbp]
\begin{center}
\includegraphics[height=2.34in,width=3.28in] {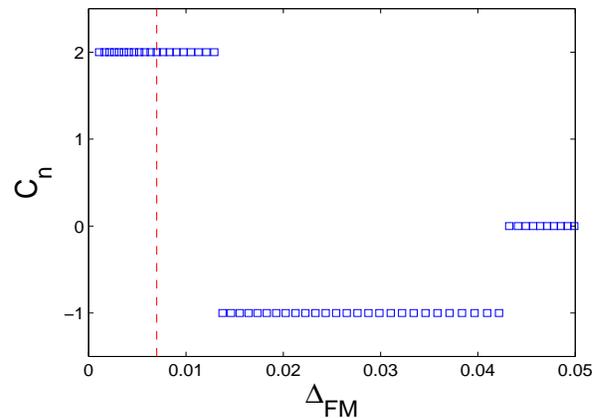}
\end{center}
\caption{(Color online) Chern number $C_{n}$ versus the FM gap $\Delta _{
\mathrm{FM}}$ at $n=1$ in the 1/4-vacancy system. Two discontinuous
transitions are at 13 meV and 43 meV, respectively. Here $V_A $ is set to be $\infty$. 
For realistic realization with a finite $V_A$, the phase boundary of $\Delta _{
\mathrm{FM}}=0$ moves to the location of the dash line so that the region between $\Delta _{
\mathrm{FM}}=0$ and the dash line becomes metallic. Here $V_A = 3t$. See the text, Sec. V,  for details.}
\label{Cn_one_forth}
\end{figure}

The FM phase diagram is shown in Fig. \ref{mz}(e) and (f). The average
moment at \textit{B} sites is defined by $m_{B}=\frac{1}{4}%
\sum\nolimits_{i=1}^{4}m_{Bi}$. Due to lattice symmetry, moments at all
\textit{A} sites are the same and moments on $B_{1}$, $B_{2}$, and $B_{3}$\
are equal. Fig. \ref{mz}(f) shows that the critical \textit{U} at half
filling is about 0.1 eV, which is larger than that of the full-vacancy
system and is close to that of the 1/2-vacancy system. However, the
ferromagnetism is the weakest compared to previous cases: the growth rate of
$m_{B}$ versus $U$ is much slower and the moment is suppressed seriously by
doping. The maximal moment will be 1/8 as indicated in Fig. \ref{mz}(f).

In spite of having weaker ferromagnetism, the 1/4-vacancy system has a
nontrivial topology at half filling and may become QAH insulator in
appropriate conditions. Fig. \ref{Cn_one_forth} shows the computed Chern
number at half filling. It is clearly shown that at half filling, the
1/4-vacancy system is a QAH state with $C_{n}=2$ when $\Delta _{\mathrm{FM}%
}\lessapprox \frac{1}{3}\lambda _{SO}$, a QAH state with $C_{n}=-1$ when $%
\frac{1}{3}\lambda _{SO}\lessapprox \Delta _{\mathrm{FM}}<\lambda _{SO}$,
and makes transition to a trivial insulator when $\Delta _{\mathrm{FM}%
}>\lambda _{SO}$. The transition occurs at the point when two midgap bands
touches and the energy gap closes. The band touch occurs when the FM gap
exceeds the original gap due to the intrinsic SOC. Hence the criterion of
the existence of nontrivial topological phase is $\Delta _{\mathrm{FM}}<\lambda _{SO}$.
By solving the above condition of band-gap closing for all midgap bands, one finds that
the transition occurs when $\Delta _{\mathrm{FM}}\approx \frac{1}{3}\lambda
_{SO}$ and $\Delta _{\mathrm{FM}}=\lambda _{SO}$. The band-touching trends
are manifested in Fig. \ref{Ek_one_fourth}: dashed lines for $\Delta _{\mathrm{%
FM}}=11$ meV and dotted one for $\Delta _{\mathrm{FM}}=38$ meV. With
increasing $\Delta _{\mathrm{FM}}$, two bands first touch at the
center of $\Gamma K^{\prime}$ when $\Delta _{\mathrm{FM}}\approx \frac{1}{3}%
\lambda _{SO}$ and then touch at $K^{\prime}$ when $\Delta _{\mathrm{FM}%
}=\lambda _{SO}$. Detailed
demonstrations of the topological transitions are relayed to Appendix A3.

The topological phases in the 1/4-vacancy system are more feasible than
those in the full-vacancy system, in which QAH state is unrealistic to be
implemented due to its narrow phase space. Here, in spite of the same
criterion $\Delta _{\mathrm{FM}}<\lambda _{SO}$ that has to be satisfied, as
shown in Fig. \ref{mz}(f), the required value of the Hubbard $U$ for $\Delta
_{\mathrm{FM}}$ to reach $\lambda _{SO}$ is much increased to 0.34 eV as
compared to 0.1 eV for the full-vacancy system. The transitions shown in
Fig. \ref{Cn_one_forth} imply that the QAH phase exists for smaller FM gaps.
As the FM gap increases, two midgap bands touch and the topology of the FM
states become trivial. Since the FM gap is proportional to the average
magnetic moment which decreases in higher temperatures, results shown in
Fig. \ref{Cn_one_forth} indicates that the associated QAH effect tends to
survive at high temperatures.
\begin{figure}[tbp]
\begin{center}
\includegraphics[width=3.28in] {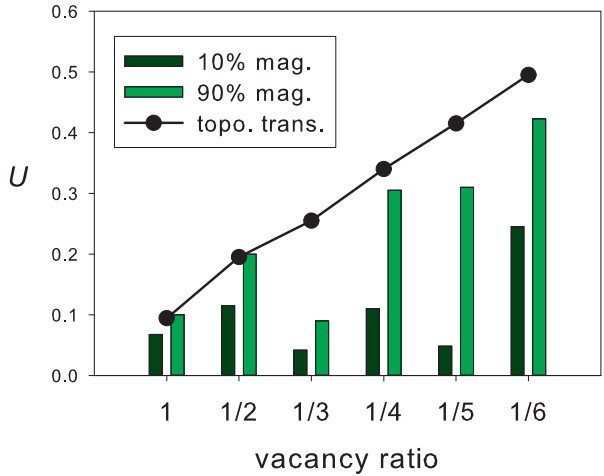}
\end{center}
\caption{(Color online) Magnitudes of $U$ (in units of eV) to reach required
magnetization in different 1/\textit{q} -vacancy systems, \textit{q} from
one to six. \textit{q} = 1 is for the full-vacancy system. Green bars are
for the \textit{U} when $m_{B}$ reaches 10\% (dark) and 90\% (light) of 1/2%
\textit{q}, respectively. The filled circles (connected by a solid line) are
the values of \textit{U} making topological phase transitions, i.e. $\Delta
_{\mathrm{FM}}=\protect\lambda _{SO}$.}
\label{Uc_comparison}
\end{figure}

\section{Summary and discussions}

In summary, we have investigated ferromagnetism induced by breaking
sublattice symmetry through saturating chemical bonds on one-side of the
buckled honeycomb lattice. The buckled geometry of germanene makes it
possible to saturate chemical bonds on one sublattice. It is shown that when
fractions of chemical bonds on one-side are saturated, two narrow midgap
bands always exist at half filling. Furthermore, the midgap bands generally
support flat band ferromagnetism in the presence of the Hubbard $U$
interaction. We have shown that given appropriate conditions, midgap bands
consist of only \textit{B}-site electrons (except zero-energy degeneracy in $%
K_{\mathbf{k}}$) and can be expressed by a spin-1/2 model.

We have considered different concentrations of chemical bonds being
saturated by setting potential to infinity at sites with saturated chemical
bonds. Specifically, by considering periodic 1/\textit{q}-vacancy
configurations, the critical interaction for ferromagnetism is highly
related to the value of \textit{q}. In Fig. \ref{Uc_comparison}, we
summarize the critical \textit{U} for ferromagnetism for different 1/\textit{%
q}-vacancy cases, \textit{q} = 1, 2, ..., 6. Two color bars for a given 1/%
\textit{q} in Fig. \ref{Uc_comparison} stand for the magnetization reaching
10\% (dark one) and 90\% (light one) of 1/2\textit{q}, respectively. As the
height difference of two is short, it means a quick magnetic saturation and
vice versa. We can see in Fig. \ref{Uc_comparison} that although the
critical \textit{U} for 10\% magnetic saturation is non-uniform, the value
of \textit{U} for 90\% saturation tends to increase as vacancy concentration
decreases (\textit{q} increases), except for the 1/3-vacancy system which
has a dissimilar band structure.

With ferromagnetism, we find that QAH states can be realized in these six 1/%
\textit{q}-vacancy systems and the condition for the QAH phase is $\Delta _{%
\mathrm{FM}}<\lambda _{SO}$. The non-trivial Chern number in these
six systems is $C_{n}=-1$ except that the 1/4-vacancy system has an
additional $C_{n}=2$ phase for weaker ferromagnetism. The reason why the
1/4-vacancy system can have a larger Chern number is that it doubles the
unit length of the vacancy-free system and preserves the $C_{3}$
symmetry. Values of \textit{U} to reach $\Delta _{\mathrm{FM}}=\lambda _{SO}$
is also shown by filled circles in Fig. \ref{Uc_comparison}. Hence the QAH phase
lies in the region below filled
circles or the solid line, in which $\Delta _{\mathrm{FM}}<\lambda _{SO}$ satisfies. 
A linear increase of the critical \textit{U} for QAH phase
is evident. Although a strong interaction leading to large ferromagnetism
that tends to break the QAH phase,  ferromagnetism
becomes weak by decreasing the vacancy ratio and the QAH phase survives
in these regimes. In the cases we studied, the
critical \textit{U} increases from 0.1 eV in full-vacancy system to 0.5 eV
in 1/6-vacancy system. The physical reason of why strong ferromagnetism breaks
the QAH phase is that in a fully polarized band, the Berry phase due to spins is suppressed. 
Therefore, in the phase when the Chern number vanishes, the corresponding spin Chern
number also vanishes. The QSH phase \cite{Ezawa2012b}
is thus excluded in our FM models.

In real implementation of saturating chemical bonds, vacancies may not be arranged in a perfect 
ordered pattern.  When vacancies are randomly distributed, the system becomes disordered. 
Empirically, it is known that disorders generally reduce the coherence of quasiparticles 
Hence bandwidths of the midgap bands get broaden.
As a result, the density of states drops and ferromagnetism is suppressed. \cite{Kogan2001}
In addition to the broadening of bandwidths, random vacancies generally introduce additional 
impurity states inside the ferromagnetic gap, which tends to deplete the
spectral weight of midgap bands that is responsible for non-trivial Berry phases.\cite{Lee2014}
As a result, as the concentration of random vacancies increases, two midgap bands
get broaden enough so that they eventually merge together with the impurity states. The ferromagnetic gap is
destroyed and the QAH effect disappears. Hence large number of random vacancies is expected to
kill the QAH effect in a similar way as disorder kills the QSH effects.\cite{Lee2014}  Therefore, as long
as concentration of random vacancies is kept low ($\le 5$\%)\cite{Lee2014}, the QAH phase can be realized.  

While in our theoretical model, the saturation of chemical bonds is modeled
by an infinity potential at given site, in experimental realization such as
realization by hydrogenation, the potential at the site with hydrogen bond
is not infinite. Hence for realistic realizations, the saturation of a chemical 
bond should be modeled by a finite $V_A$ with relevant hopping amplitudes
($t'$) to the saturated site being reduced. For finite potential $V_{A}$, an effective hopping of $
t'^{2}/V_{A} $ between \textit{B} sites will be induced so that the bandwidth
of the impurity band would be larger. This would increase the critical
interaction $U_{c}$ for ferromagnetism to occur. A larger $U_{c}$ implies a
larger FM gap, which is confirmed by the first-principles calculation. \cite
{XQWang2012} However, due to topological robustness of QAH states, we expect
that the Berry curvature of the midgap bands and the corresponding Chern
number would remain the same. The effects due to finite $V_A$ is examined
in Fig.~\ref{Cn_one_forth}. Here we find that for
realistic realization with a finite $V_A$, the phase boundary of $\Delta _{
\mathrm{FM}}=0$ moves to the location of the dash line so that the region between $\Delta _{
\mathrm{FM}}=0$ and the dash line becomes metallic. Here $V_A = 3t$
and $t'=0.1t$. For smaller $t'=0.05t$, the metallic phase vanishes. 
For general finite $V_A$,  in order for the system to be an insulator, 
the FM gap needs be larger than the bandwidth due to the effective hopping.
Hence one requires $\Delta _{\mathrm{FM}}>4t'^{2}/V_{A}$. 
Since the QAH phase requires $\Delta _{\mathrm{FM}} <\lambda _{SO}$, 
it leads to the condition that preserves the QAH phases: $\lambda_{SO}>4t'^{2}/V_{A}$.  
For typical magnitudes of $V_A$,  we find that when $V_A=5t$, $t'<0.2t$ and 
when $V_A=3t$, $t'<0.16t$, which is consistent with results found in Fig.~\ref{Cn_one_forth}.
Since conditions on hopping amplitudes found for typical $V_A$ are in the typical range of  
hopping amplitudes, the above analysis indicates 
that it is feasible to realize the proposed QAH states by saturating chemical bonds.

In addition to the above-mentioned issue of magnitude of the potential, in
real situation, there are also potential complications due to local
distortion of the honey lattice. For instance, the appearance of vacancy may
modify electronic parameters locally. The perturbation may also lift up the $%
\sigma $ bands at $\Gamma $, \cite{Lew2010,Houssa2011,PZhang2012,XQWang2012}
which may change the topology. However, a recent finding shows that
hydrogenated graphene actually enhances SOC, \cite{Balakrishnan2013} which
can further stabilize the proposed QAH states. It is also shown that by
using appropriate substrate, \cite{HSLiu2013} the $\pi $ bands can be
preserved above the $\sigma $ bands. Although our work does not include all
these detailed complications, our results indicate that spontaneous QAH
states tend to occur in low concentration limit of vacancies. In low
concentration of vacancies, effects due to local distortions are diluted and
one expects that low energy physics at half filling is dominated by midgap
bands induced by vacancies. Our results thus provide a potential way to
engineer buckled honeycomb structures into high-temperature QAH insulators.

\begin{acknowledgments}
The work was supported by the National Science Council of Taiwan. We would
like to thank Prof. Ming-Che Chang and J. S. You for helpful discussions.
\end{acknowledgments}

\appendix{}

\section{Band structures for the 1/2, 1/3, and 1/4-vacancy systems}

\subsection{1/2-vacancy}

According to the lattice shown in Fig. \ref{lattice}(a), the bulk
Hamiltonian for the 1/2-vacancy system can be written as $%
H=H_{t}+H_{SO}+H_{R2}+H_{FM}$ with
\begin{equation}
H_{t}=\sum_{\mathbf{k},\sigma }\Psi _{\sigma }^{\dag }(\mathbf{k})\left(
\begin{array}{ccc}
0 & \gamma _{0}(\mathbf{k}) & -t \\
\gamma _{0}^{\ast }(\mathbf{k}) & 0 & 0 \\
-t & 0 & 0%
\end{array}%
\right) \Psi _{\sigma }(\mathbf{k})  \label{Ht2}
\end{equation}%
\begin{widetext}%
\begin{equation}
H_{SO}=\sum_{\mathbf{k},\sigma }\sigma \Psi _{\sigma }^{\dag }(%
\mathbf{k})\left(
\begin{array}{ccc}
\gamma _{1}(\mathbf{k}) & 0 & 0 \\
0 & -\gamma _{1}(\mathbf{k}) & f^{\ast }(\mathbf{k})\gamma _{2}(\mathbf{k})
\\
0 & f(\mathbf{k})\gamma _{2}(\mathbf{k}) & -\gamma _{1}(\mathbf{k})%
\end{array}%
\right) \Psi _{\sigma }(\mathbf{k})
\end{equation}%
\begin{equation}
H_{R2}=\sum_{\mathbf{k}}\Psi _{\uparrow }^{\dag }(\mathbf{k}%
)\left(
\begin{array}{ccc}
-i\gamma _{3}(\mathbf{k}) & 0 & 0 \\
0 & i\gamma _{3}(\mathbf{k}) & f^{\ast }(\mathbf{k})\gamma _{4}(\mathbf{k})
\\
0 & f(\mathbf{k})\gamma _{4}(\mathbf{k}) & i\gamma _{3}(\mathbf{k})%
\end{array}%
\right) \Psi _{\downarrow }(\mathbf{k})+H.c.
\end{equation}%
\end{widetext}%
\begin{equation}
H_{FM}=-\sum_{\mathbf{k},\sigma }\sigma \Psi _{\sigma }^{\dag }(\mathbf{k}%
)\left(
\begin{array}{ccc}
\Delta _{A} & 0 & 0 \\
0 & \Delta _{B1} & 0 \\
0 & 0 & \Delta _{B2}%
\end{array}%
\right) \Psi _{\sigma }(\mathbf{k}),
\end{equation}%
where $\Psi _{\sigma }^{\dag }(\mathbf{k})=(c_{A,\sigma }^{\dag }(\mathbf{k}%
),c_{B1,\sigma }^{\dag }(\mathbf{k}),c_{B2,\sigma }^{\dag }(\mathbf{k}))$, $%
\gamma _{0}(\mathbf{k})=-2t\cos (\frac{1}{2}k_{y}a)e^{ik_{y}a/2}$, $\gamma
_{1}(\mathbf{k})=\frac{2\lambda _{SO}}{3\sqrt{3}}\sin \left( k_{y}a\right) $%
, $\gamma _{2}(\mathbf{k})=\frac{4\lambda _{SO}}{3\sqrt{3}}\cos (\frac{1}{2}%
k_{x}b)\sin \left( \frac{1}{2}k_{y}a\right) $, $\gamma _{3}(\mathbf{k})=%
\frac{4}{3}\lambda _{R2}\sin (k_{y}a)$, $\gamma _{4}(\mathbf{k})=-\frac{4}{3}%
\lambda _{R2}\left[ \sqrt{3}\sin (\frac{1}{2}k_{x}b)\cos \left( \frac{1}{2}%
k_{y}a\right) -i\cos (\frac{1}{2}k_{x}b)\sin \left( \frac{1}{2}k_{y}a\right) %
\right] $, and $f(\mathbf{k})=e^{ik_{x}b/2}e^{ik_{y}a/2}$. The unit length
in the \textit{x} direction is $b=\sqrt{3}a$. FM gaps are defined by $\Delta
_{A}=Um_{A}$ and $\Delta _{Bi}=Um_{Bi}$ for$\mathit{\ }i=1,2$.

The midgap bands can be constructed explicitly. For the hopping Hamiltonian $%
H_{t}$ in Eq. (\ref{Ht2}), the eigenstate with eigenvalue zero is given by
\begin{equation*}
\left\vert 0\right\rangle =\left(
\begin{array}{c}
u_{1} \\
u_{2}%
\end{array}%
\right) =\frac{1}{D}\left(
\begin{array}{c}
e^{-ik_{y}a/2} \\
-2\cos \left( \frac{1}{2}k_{y}a\right)%
\end{array}%
\right) ,
\end{equation*}%
where \textit{D} is a normalization factor and is given by $D=\sqrt{1+4\cos
^{2}\left( \frac{1}{2}k_{y}a\right) }$. Following Eq. (\ref{Hmid}), using
this basis, the effective Hamiltonian for midgap bands of the 1/2-vacancy
system is characterized by expectation values of SOC and the FM field, which
are
\begin{eqnarray*}
m_{\mathbf{k}} &=&-\Delta _{B1}\left\vert u_{1}\right\vert ^{2}-\Delta
_{B2}\left\vert u_{2}\right\vert ^{2}-\gamma _{1}+\left( W+W^{\ast }\right)
\gamma _{2}, \\
\delta _{\mathbf{k}} &=&i\gamma _{3}+\left( W+W^{\ast }\right) \gamma _{4},
\end{eqnarray*}%
where $W=f^{\ast }u_{1}^{\ast }u_{2}$.

The topological phase transition occurs at band-touching when the
corresponding gap closes. At half filling, band-touching occurs between two
midgap bands. Therefore, we shall need to find the \textbf{k} point(s) at
which the gap between two midgap bands closes. In Eq. (\ref{Hmid}), the gap
closes when both $m_{\mathbf{k}}$ and $\delta _{\mathbf{k}}$ vanish. It can
be identified the \textbf{k} point will be $\mathbf{k}=\left( 0,-\frac{2\pi
}{3a}\right) $ at which $\delta _{\mathbf{k}}=0$ and at the same time $m_{%
\mathbf{k}}=-\frac{1}{2}\left( \Delta _{B1}+\Delta _{B2}\right) +\lambda
_{SO}=-\Delta _{\mathrm{FM}}+\lambda _{SO}$ (refer to Fig. \ref{Ek_one_half}%
). As the result, the topological phase transition happens at $\Delta _{%
\mathrm{FM}}=\lambda _{SO}$.

\subsection{1/3-vacancy}

According to the lattice in Fig. \ref{lattice}(b), the bulk Hamiltonian for
the 1/3-vacancy system is $H=H_{t}+H_{SO}+H_{R2}+H_{FM}$ with%
\begin{equation}
H_{t}=-t\sum_{\mathbf{k},\sigma }\Psi _{\sigma }^{\dag }(\mathbf{k})\left(
\begin{array}{ccccc}
0 & 0 & 1 & 1 & 1 \\
0 & 0 & 1 & f_{1}(\mathbf{k}) & f_{2}(\mathbf{k}) \\
1 & 1 & 0 & 0 & 0 \\
1 & f_{1}^{\ast }(\mathbf{k}) & 0 & 0 & 0 \\
1 & f_{2}^{\ast }(\mathbf{k}) & 0 & 0 & 0%
\end{array}%
\right) \Psi _{\sigma }(\mathbf{k}),  \label{Ht3}
\end{equation}%
\begin{equation}
H_{SO}=\sum_{\mathbf{k},\sigma }\sigma \Psi _{\sigma }^{\dag }(\mathbf{k}%
)\left(
\begin{array}{cc}
M_{A}(\mathbf{k}) & \mathbf{0}_{2\times 3} \\
\mathbf{0}_{3\times 2} & M_{B}(\mathbf{k})%
\end{array}%
\right) \Psi _{\sigma }(\mathbf{k})
\end{equation}%
\begin{equation}
H_{R2}=\sum_{\mathbf{k}}\Psi _{\uparrow }^{\dag }(\mathbf{k})\left(
\begin{array}{cc}
\Delta _{A}^{\prime }(\mathbf{k}) & \mathbf{0}_{2\times 3} \\
\mathbf{0}_{3\times 2} & \Delta _{B}^{\prime }(\mathbf{k})%
\end{array}%
\right) \Psi _{\downarrow }(\mathbf{k})+H.c.
\end{equation}%
\begin{equation}
H_{FM}=-\sum_{\mathbf{k},\sigma }\sigma \Psi _{\sigma }^{\dag }(\mathbf{k})%
\text{diag}\left(
\begin{array}{c}
\Delta _{A} \\
\Delta _{A} \\
\Delta _{B} \\
\Delta _{B} \\
\Delta _{B}%
\end{array}%
\right) \Psi _{\sigma }(\mathbf{k}),
\end{equation}%
with
\begin{equation*}
M_{A}(\mathbf{k})=\frac{\lambda _{SO}}{3\sqrt{3}}\left(
\begin{array}{cc}
0 & g_{0}^{\ast }(\mathbf{k}) \\
g_{0}(\mathbf{k}) & 0%
\end{array}%
\right) ,
\end{equation*}%
\begin{equation*}
M_{B}(\mathbf{k})=-\frac{\lambda _{SO}}{3\sqrt{3}}\left(
\begin{array}{ccc}
0 & f_{1}(\mathbf{k})g_{0}^{\ast }(\mathbf{k}) & g_{0}(\mathbf{k}) \\
f_{1}^{\ast }(\mathbf{k})g_{0}(\mathbf{k}) & 0 & f_{2}(\mathbf{k}%
)g_{0}^{\ast }(\mathbf{k}) \\
g_{0}^{\ast }(\mathbf{k}) & f_{2}^{\ast }(\mathbf{k})g_{0}(\mathbf{k}) & 0%
\end{array}%
\right) ,
\end{equation*}%
\begin{equation*}
\Delta _{A}(\mathbf{k})=\frac{2}{3}\lambda _{R2}\left(
\begin{array}{cc}
0 & g_{2}(\mathbf{k}) \\
g_{1}(\mathbf{k}) & 0%
\end{array}%
\right) ,
\end{equation*}%
\begin{equation*}
\Delta _{B}(\mathbf{k})=-\frac{2}{3}\lambda _{R2}\left(
\begin{array}{ccc}
0 & f_{1}(\mathbf{k})g_{2}(\mathbf{k}) & g_{1}(\mathbf{k}) \\
f_{1}^{\ast }(\mathbf{k})g_{1}(\mathbf{k}) & 0 & f_{2}(\mathbf{k})g_{2}(%
\mathbf{k}) \\
g_{2}(\mathbf{k}) & f_{2}^{\ast }(\mathbf{k})g_{1}(\mathbf{k}) & 0%
\end{array}%
\right) ,
\end{equation*}%
where $\Psi _{\sigma }^{\dag }(\mathbf{k})=(\Psi _{A\sigma }^{\dag }(\mathbf{%
k}),\Psi _{B\sigma }^{\dag }(\mathbf{k}))$ with $\Psi _{A\sigma }^{\dag }(%
\mathbf{k})=(c_{A1,\sigma }^{\dag }(\mathbf{k}),c_{A2,\sigma }^{\dag }(%
\mathbf{k}))$ and $\Psi _{B\sigma }^{\dag }(\mathbf{k})=(c_{B1,\sigma
}^{\dag }(\mathbf{k}),c_{B2,\sigma }^{\dag }(\mathbf{k}),c_{B3,\sigma
}^{\dag }(\mathbf{k}))$. Here relevant functions are given by $g_{0}(\mathbf{%
k})=i\left[ 1+f_{1}(\mathbf{k})+f_{2}(\mathbf{k})\right] $, $g_{1}(\mathbf{k}%
)=\left[ 1-e^{i\frac{\pi }{3}}f_{1}(\mathbf{k})-e^{-i\frac{\pi }{3}}f_{2}(%
\mathbf{k})\right] $, $g_{2}(\mathbf{k})=-\left[ 1-e^{i\frac{\pi }{3}%
}f_{1}^{\ast }(\mathbf{k})-e^{-i\frac{\pi }{3}}f_{2}^{\ast }(\mathbf{k})%
\right] $, and $f_{i}(\mathbf{k})=e^{i\mathbf{k\cdot }\vec{b}_{i}}$ $(i=1,2)$
for $\vec{b}_{1}=b\hat{x}$, $\vec{b}_{2}=b(\frac{1}{2}\hat{x}+\frac{\sqrt{3}%
}{2}\hat{y})$, $\vec{b}_{3}=b(\frac{1}{2}\hat{x}-\frac{\sqrt{3}}{2}\hat{y})$
with $b=\sqrt{3}a$. Due to symmetry, FM gaps on two \textit{A} sites and
those on \textit{B} sites are respectively the same.

For the midgap band, the zero-energy eigenstate $\left\vert 0\right\rangle $
of the hopping Hamiltonian $H_{t}$ in Eq. (\ref{Ht3}) is
\begin{equation*}
\left\vert 0\right\rangle =\left(
\begin{array}{c}
u_{1} \\
u_{2} \\
u_{3}%
\end{array}%
\right) =\frac{1}{D}\left(
\begin{array}{c}
f_{1}-f_{2} \\
f_{2}-1 \\
1-f_{1}%
\end{array}%
\right)
\end{equation*}%
with the normalization factor $D=\sqrt{6-2\textstyle\sum\nolimits_{i=1}^{3}%
\cos \mathbf{k\cdot }\vec{b}_{i}}$. By using this wavefunction, $m_{\mathbf{k%
}}$ and $\delta _{\mathbf{k}}$ for the effective Hamiltonian of midgap bands
are given by
\begin{eqnarray*}
m_{\mathbf{k}} &=&-\Delta _{\mathrm{FM}}-\left[ \gamma _{SO}W^{\prime }+H.c.%
\right] , \\
\delta _{\mathbf{k}} &=&\gamma _{1}W^{\prime }+\gamma _{2}W^{\prime \ast },
\end{eqnarray*}%
where $W^{\prime }=f_{1}^{\ast }u_{2}^{\ast }u_{1}+f_{2}^{\ast }u_{3}^{\ast
}u_{2}+u_{1}^{\ast }u_{3}$, $\gamma _{SO}=\frac{1}{3\sqrt{3}}\lambda
_{SO}g_{0}$, $\gamma _{1}=-\frac{2}{3}\lambda _{R2}g_{1}$, and $\gamma _{2}=%
\frac{2}{3}\lambda _{R2}g_{2}$. However, the choice of $\left\vert
0\right\rangle $ at $\Gamma $ is singular, so we have to choose another
gauge. This is due to non-uniqueness of the zero-energy eigenstate. At $%
\Gamma $ there is degeneracy of three\ and three wavefunctions are $\psi
_{1}=\frac{1}{\sqrt{2}}\left( 1,-1,0,0,0\right) ^{t}$, $\psi _{2}=\frac{1}{%
\sqrt{6}}\left( 0,0,2,-1,-1\right) ^{t}$, and $\psi _{3}=\frac{1}{\sqrt{2}}%
\left( 0,0,0,1,-1\right) ^{t}$, respectively. The first one is constructed
by \textit{A}-site electrons while the last two are by \textit{B}-site
electrons. Now the midgap band will contain \textit{A}-site electrons though
a small proportion. This degeneracy explains samll gaps (of $\lambda _{SO}$)
opened between the midgap band and its neighbors in Fig. \ref{dos}(c).

Finally, we identify the transition point. The band-touching occurs at $
\mathbf{k}=\Gamma $, where $\Delta _{A}(\mathbf{k})=\Delta _{B}(\mathbf{k}%
)=0$ so that spins are decoupled. The effective Hamiltonian of a given spin $
\sigma $ for these three states $\psi _{1,2,3}$ is
\begin{equation*}
\mathcal{H}_{\Gamma ,\sigma }=-\sigma \left(
\begin{array}{ccc}
\Delta _{A} & 0 & 0 \\
0 & \Delta _{B} & i\lambda _{SO} \\
0 & -i\lambda _{SO} & \Delta _{B}%
\end{array}%
\right).
\end{equation*}%
We find that energy eigenvalues are $-\sigma \Delta _{A}$, $-\sigma \left( \Delta
_{B}\pm \lambda _{SO}\right) $ and hence the condition for band touching is $\Delta
_{B}=\lambda _{SO}$ ($\lambda _{SO}$, $\Delta _{B}>0$) as we expect. Here there is a finite 
(but tiny) $\Delta _{A}$ so that the system is insulating. Note that this tiny $\Delta _{A}$ is present though
it is difficult to be identified in Fig.~\ref{Ek_one_third}.

\subsection{1/4-vacancy}

According to the lattice given by Fig. \ref{lattice}(c), the bulk
Hamiltonian for the 1/4-vacancy system is written by $%
H=H_{t}+H_{SO}+H_{R2}+H_{FM}$ where

\begin{equation}
H_{t}=\sum_{\mathbf{k},\sigma }\Psi _{\sigma }^{\dag }(\mathbf{k})\left(
\begin{array}{cc}
\mathbf{0}_{3\times 3} & T(\mathbf{k}) \\
T^{\dag }(\mathbf{k}) & \mathbf{0}_{4\times 4}%
\end{array}%
\right) T(\mathbf{k})\Psi _{\sigma }(\mathbf{k})  \label{Ht4}
\end{equation}%
\begin{equation}
H_{SO}=\sum_{\mathbf{k},\sigma }\sigma \Psi _{\sigma }^{\dag }(\mathbf{k}%
)\left(
\begin{array}{cc}
M_{A}^{\prime }(\mathbf{k}) & \mathbf{0}_{3\times 4} \\
\mathbf{0}_{4\times 3} & M_{B}^{\prime }(\mathbf{k})%
\end{array}%
\right) \Psi _{\sigma }(\mathbf{k})
\end{equation}%
\begin{equation}
H_{R2}=\sum_{\mathbf{k},\sigma }\Psi _{\uparrow }^{\dag }(\mathbf{k})\left(
\begin{array}{cc}
\Delta _{A}^{\prime }(\mathbf{k}) & \mathbf{0}_{3\times 4} \\
\mathbf{0}_{4\times 3} & \Delta _{B}^{\prime }(\mathbf{k})%
\end{array}%
\right) \Psi _{\downarrow }(\mathbf{k})+H.c.
\end{equation}%
\begin{equation}
H_{FM}=-\sum_{\mathbf{k},\sigma }\sigma \Psi _{\sigma }^{\dag }(\mathbf{k})%
\text{diag}\left(
\begin{array}{c}
\Delta _{A1} \\
\Delta _{A2} \\
\Delta _{A3} \\
\Delta _{B1} \\
\Delta _{B2} \\
\Delta _{B3} \\
\Delta _{B4}%
\end{array}%
\right) \Psi _{\sigma }(\mathbf{k}),
\end{equation}%
with%
\begin{equation*}
T(\mathbf{k})=-t\left(
\begin{array}{cccc}
0 & 1 & 1 & 1 \\
f_{1}^{\prime \ast }(\mathbf{k}) & f_{1}^{\prime \ast }(\mathbf{k}) & 0 & 1
\\
f_{2}^{\prime \ast }(\mathbf{k}) & 0 & f_{2}^{\ast }(\mathbf{k}) & 1%
\end{array}%
\right) ,
\end{equation*}%
\begin{equation*}
M_{A}^{\prime }(\mathbf{k})=\frac{1}{3\sqrt{3}}\lambda _{SO}\left(
\begin{array}{ccc}
0 & -s_{1}(\mathbf{k}) & s_{2}(\mathbf{k}) \\
-s_{1}^{\ast }(\mathbf{k}) & 0 & s_{3}^{\ast }(\mathbf{k}) \\
s_{2}^{\ast }(\mathbf{k}) & s_{3}(\mathbf{k}) & 0%
\end{array}%
\right) ,
\end{equation*}%
\begin{equation*}
\Delta _{A}^{\prime }(\mathbf{k})=\frac{2}{3}\lambda _{R2}\left(
\begin{array}{ccc}
0 & e^{-i\frac{\pi }{6}}s_{1}(\mathbf{k}) & -is_{2}(\mathbf{k}) \\
e^{-i\frac{\pi }{6}}s_{1}^{\ast }(\mathbf{k}) & 0 & e^{i\frac{\pi }{6}%
}s_{3}^{\ast }(\mathbf{k}) \\
-is_{2}^{\ast }(\mathbf{k}) & e^{i\frac{\pi }{6}}s_{3}(\mathbf{k}) & 0%
\end{array}%
\right) ,
\end{equation*}%
and
\begin{widetext}%
\begin{equation*}
M_{B}^{\prime }(\mathbf{k})=-\frac{1}{3\sqrt{3}}\lambda _{SO}\left(
\begin{array}{cccc}
0 & s_{2}(\mathbf{k}) & -s_{1}(\mathbf{k}) & f_{1}^{\prime }(\mathbf{k}%
)s_{3}^{\ast }(\mathbf{k}) \\
s_{2}^{\ast }(\mathbf{k}) & 0 & s_{3}(\mathbf{k}) & -s_{1}(\mathbf{k}) \\
-s_{1}^{\ast }(\mathbf{k}) & s_{3}^{\ast }(\mathbf{k}) & 0 & s_{2}(\mathbf{k}%
) \\
f_{1}^{\prime \ast }(\mathbf{k})s_{3}(\mathbf{k}) & -s_{1}^{\ast }(\mathbf{k}%
) & s_{2}^{\ast }(\mathbf{k}) & 0%
\end{array}%
\right) ,
\end{equation*}%
\begin{equation*}
\Delta_{B}^{\prime }(\mathbf{k})=-\frac{2}{3}\lambda _{R2}\left(
\begin{array}{cccc}
0 & -i s_{2}(\mathbf{k}) & e^{-i\frac{\pi }{6}}s_{1}(\mathbf{k}) & e^{i%
\frac{\pi }{6}}f_{1}^{\prime }(\mathbf{k})s_{3}^{\ast }(\mathbf{k}) \\
-i s_{2}^{\ast }(\mathbf{k}) & 0 & e^{i\frac{\pi }{6}}s_{3}(%
\mathbf{k}) & e^{-i\frac{\pi }{6}}s_{1}(\mathbf{k}) \\
e^{-i\frac{\pi }{6}}s_{1}^{\ast }(\mathbf{k}) & e^{i\frac{\pi
}{6}}s_{3}^{\ast }(\mathbf{k}) & 0 & -i s_{2}(\mathbf{k}) \\
e^{i\frac{\pi }{6}}f_{1}^{\prime \ast }(\mathbf{k})s_{3}(\mathbf{k}) & e^{-i%
\frac{\pi }{6}}s_{1}^{\ast }(\mathbf{k}) & -i s_{2}^{\ast }(%
\mathbf{k}) & 0%
\end{array}%
\right) .
\end{equation*}%
\end{widetext}Here the basis is $\Psi _{\sigma }^{\dag }(\mathbf{k})=(\Psi
_{A\sigma }^{\dag }(\mathbf{k}),\Psi _{B\sigma }^{\dag }(\mathbf{k}))$ with $%
\Psi _{A\sigma }^{\dag }(\mathbf{k})=(c_{A1,\sigma }^{\dag }(\mathbf{k}%
),c_{A2,\sigma }^{\dag }(\mathbf{k}),c_{A3,\sigma }^{\dag }(\mathbf{k}))$
and $\Psi _{B\sigma }^{\dag }(\mathbf{k})=(c_{B1,\sigma }^{\dag }(\mathbf{k}%
),c_{B2,\sigma }^{\dag }(\mathbf{k}),c_{B3,\sigma }^{\dag }(\mathbf{k}%
),c_{B4,\sigma }^{\dag }(\mathbf{k}))$. Relevant functions are given by $%
f_{i}^{\prime }(\mathbf{k})=e^{i\mathbf{k\cdot }\vec{c}_{i}}$ and $s_{i}(%
\mathbf{k})=2e^{i\mathbf{k\cdot }\vec{c}_{i}/2}\sin \left( \frac{1}{2}%
\mathbf{k\cdot }\vec{c}_{i}\right) $ for $i=1,2,3$. Here $\vec{c}_{1}=c(%
\frac{\sqrt{3}}{2}\hat{x}+\frac{1}{2}\hat{y})$, $\vec{c}_{2}=\hat{y}c$, and $%
\vec{c}_{3}=c(\frac{\sqrt{3}}{2}\hat{x}-\frac{1}{2}\hat{y})$ with $c=2a$.
The FM gaps are defined by $\Delta _{Ai}=Um_{Ai}$ $(i=1,2,3)$ and $\Delta
_{Bi}=Um_{Bi}$ $(i=1,2,3,4)$.

The zero-energy eigenstate $\left\vert 0\right\rangle $ of $H_{t}$ in Eq. (%
\ref{Ht4}) is given by
\begin{equation*}
\left\vert 0\right\rangle =\left(
\begin{array}{c}
u_{1} \\
u_{2} \\
u_{3} \\
u_{4}%
\end{array}%
\right) =\frac{1}{D}\left(
\begin{array}{c}
1-f_{1}^{\prime }-f_{2}^{\prime } \\
-1-f_{1}^{\prime }+f_{2}^{\prime } \\
-1+f_{1}^{\prime }-f_{2}^{\prime } \\
2%
\end{array}%
\right)
\end{equation*}%
with a normalization $D=\sqrt{13-2\textstyle\sum\nolimits_{i=1}^{3}\cos
\mathbf{k\cdot }\vec{c}_{i}}$. Using the wavefunction as basis, the
effective Hamiltonian for the midgap bands are characterized by $m_{\mathbf{k%
}}$ and $\delta _{\mathbf{k}}$, which are given by
\begin{eqnarray*}
m_{\mathbf{k}} &=&-\textstyle\sum\nolimits_{i=1}^{4}\Delta _{Bi}\left\vert
u_{i}\right\vert ^{2}-\frac{2}{3\sqrt{3}}\lambda _{SO}\Re \left(
W_{1}+W_{2}+W_{3}\right) , \\
\delta _{\mathbf{k}} &=&-\frac{4}{3}e^{-i\frac{\pi }{6}}\lambda _{R2}\left(
\Re W_{1}+e^{i\frac{2\pi }{3}}\Re W_{2}+e^{i\frac{4\pi }{3}}\Re W_{3}\right)
,
\end{eqnarray*}%
where $W_{1}=-s_{1}\left( u_{1}^{\ast }u_{3}+u_{2}^{\ast }u_{4}\right) $, $%
W_{2}=s_{2}\left( u_{1}^{\ast }u_{2}+u_{3}^{\ast }u_{4}\right) $, $%
W_{3}=s_{3}\left( u_{2}^{\ast }u_{3}+f_{1}^{\prime \ast }u_{4}^{\ast
}u_{1}\right) $.

Finally, we examine the topological phase transitions. The gap closes when
both $m_{\mathbf{k}}$ and $\delta _{\mathbf{k}}$ vanish. There are twelve
points at which $\delta _{\mathbf{k}}=0$. These points are $\Gamma $, three $%
M$s, $\pm K$, and stars of $\mathbf{k}=\pm \left( 0,\frac{2\pi }{3c}\right)
\equiv \pm \frac{K}{2}$, respectively. \cite{note1} When $\Delta _{\mathrm{FM%
}}=0$, $\Gamma $ and $M$s are Dirac points. Since at these points $\Re
W_{i}=0$ $(i=1,2,3)$, one gets $m_{\mathbf{k}}=0$ when $\Delta _{\mathrm{FM}%
}=0$. For finite $\Delta _{\mathrm{FM}}$, however, gaps are determined by $%
\pm K=\pm \left( 0,\frac{4\pi }{3c}\right) $ and $\pm \frac{K}{2}$. Hence
vanishing of gaps at these points determine the two topological phase
transitions as shown in Fig. \ref{Cn_one_forth}. Since at $\mathbf{k}=\frac{1%
}{2}K$ one finds that $m_{\mathbf{k}}=-\frac{1}{3}\left( \Delta _{B1}+\Delta
_{B2}+\Delta _{B4}\right) \mp \frac{1}{3}\lambda _{SO}\approx -\Delta _{%
\mathrm{FM}}\mp \frac{1}{3}\lambda _{SO}$, the first band touch will happen
at star of $-\frac{K}{2}$ when $\Delta _{\mathrm{FM}}\approx \frac{1}{3}%
\lambda _{SO}$. Moreover, since at $\mathbf{k}=\pm K$ one obtains $m_{%
\mathbf{k}}=-\Delta _{\mathrm{FM}}\mp \lambda _{SO}$ ($\Delta _{\mathrm{FM}}=%
\frac{1}{4}\sum\nolimits_{i=1}^{4}\Delta _{Bi}$), the second band touching
will happen at $-K$ when $\Delta _{\mathrm{FM}}=\lambda _{SO}$. (Readers can
refer to Fig. \ref{Ek_one_fourth} for the band evolution with $\Delta _{%
\mathrm{FM}}$.) These two bands touch at $\Delta _{\mathrm{FM}}\approx
\frac{1}{3}\lambda _{SO}$ and $\Delta _{\mathrm{FM}}=\lambda _{SO}$, which explains
results shown in Fig. \ref{Cn_one_forth}. In addition, for the first transition
because of band-touching at three points, the Chern number changes by three, going
from $C_{n}=2$ to $C_{n}=-1$.

\end{document}